\documentclass
[pra,onecolumn,letterpaper,noshowpacs,preprint,12pt,titlepage]{revtex4}%
\usepackage{amsfonts}
\usepackage{amsmath}
\usepackage{amssymb}
\usepackage{graphicx}%
\begin{document}
\title{Efficient Implementation of Controlled-Rotations by Using Entanglement}
\author{Ming-Yong Ye}
\author{Yong-Sheng Zhang}
\email{yshzhang@ustc.edu.cn}
\author{Guang-Can Guo}
\email{gcguo@ustc.edu.cn}
\affiliation{Key Laboratory of Quantum Information, Department of Physics, University of
Science and Technology of China (CAS), Hefei 230026, People's Republic of China}

\begin{abstract}
Implementation of controlled-rotations using entanglement is considered. We
show that the successful probability is closely related to the entanglement
and the rotation angle. The successful probability will increase if we
increase the entanglement we use or decrease the controlled-rotation angle and
the probability will trend to unit when the entangled state trends to a Bell
state or the controlled-rotation angle trends to zero.

PACS number(s): 03.67.Mn, 03.67.Ud, 03.67.Ta

\end{abstract}
\maketitle

Quantum entanglement has no classical analog and such a novel characteristic
leads to many applications in quantum information science. In Bennett
\textit{et al}'s famous quantum teleportation protocol, entanglement is used
as a channel to transmit quantum state from one qubit to another spatially
separated qubit \cite{bennett}. Quantum entanglement as an useful resource can
be stored, transmitted, manipulated and it can be created by application of
nonlocal gate on an initial non-entangled state \cite{book}. The relation
between entanglement and nonlocal gates has been widely investigated
\cite{cirac,dur,vidal,kraus,leifer,ye}. Among these how entanglement can be
used to implement nonlocal gates on spatially separated qubits is very
important, since distributed quantum computation usually needs such an
application of entanglement \cite{eisert,reznik}.

It has been shown that a maximally entangled state (Bell state) can be used to
implement any controlled-gates with unit probability if local operations and
classical communication (LOCC) are permitted \cite{eisert,reznik}. Using
partially entangled states to implement controlled-gates has also been
discussed \cite{groisman,zheng,chen}. One trivial method is transforming
partially entangled states into Bell states first, and then using the created
Bell states to implement desired gates. The successful probability of this
method relies on the entanglement transformation in the first step. This
method is not efficient especially when the resource states we use have small
entanglement. Groisman \textit{et al} have given another way to implement
controlled-gates using partially entangled states that can reach a higher
successful probability \cite{groisman}, and recently a more efficient protocol
is proposed by Chen \textit{et al} \cite{chen}.

Finding a more efficient way to implement nonlocal gates using entanglement is
the goal in this research direction. The successful probability is closely
related to the efficient of distributed quantum computers. Though the
successful probability can not reach unit when partially entangled states are
used, while Duan and Raussendorf recently show that even a small successful
probability can do efficient quantum computation through a cluster state
method \cite{duan}. It has been proven that a two-qubit entangled state can
only implement a controlled-rotation \cite{dur}. We think the following
requirements are reasonable for an efficient method that using entanglement to
implement controlled-rotations: (1) If we fix the controlled-rotation angle,
the more entanglement the state has, the bigger the successful probability we
can achieve, and when the entangled state trends to a maximally entangled
state (Bell state), the successful probability trends to unit. (2) If we fix
the entangled state we use, the smaller the controlled-rotation angle we want
to implement, the bigger the successful probability we can achieve, and when
the controlled-rotation angle trends to zero, the successful probability
trends to unit. In this paper, we propose a method to implement nonlocal gate
on two spatially separated qubits using entanglement that satisfies the above
two requirements for efficient protocols. And we show that when the
controlled-rotation angle smaller than one certain value the gate can be
implemented determinately using less than one ebit entanglement on average.

Without loss of generality we suppose Alice and Bob share the following
entangled state%
\begin{equation}
\left\vert \Psi_{ab}\right\rangle =\cos\left(  \alpha/2\right)  \left\vert
0\right\rangle _{a}\left\vert 0\right\rangle _{b}+i\sin\left(  \alpha
/2\right)  \left\vert 1\right\rangle _{a}\left\vert 1\right\rangle _{b}%
,\alpha\in\left(  0,\pi/2\right)  . \label{1}%
\end{equation}
Since this entangled state can only be used to implement controlled-rotations
we assume they want to implement the gate
\begin{equation}
U\left(  \theta\right)  =\cos\left(  \frac{\theta}{2}\right)  I_{4}%
+i\sin\left(  \frac{\theta}{2}\right)  \sigma_{z}^{A}\sigma_{z}^{B}%
=e^{i\frac{\theta}{2}\sigma_{z}^{A}\sigma_{z}^{B}},\theta\in\left(
0,\pi/2\right]  \label{2}%
\end{equation}
on qubits $A$ and $B$, where qubit $A$ and $a$ on Alice's side, $B$ and $b$ on
Bob's side. Alice and Bob's strategy is divided into the following steps:

(1) Alice implements a controlled-phase gate
\begin{equation}
U_{aA}=\left\vert 0\right\rangle _{a}\left\langle 0\right\vert _{a}\otimes
I_{2}+\left\vert 1\right\rangle _{a}\left\langle 1\right\vert _{a}%
\otimes\sigma_{z}^{A} \label{3}%
\end{equation}
on qubit $a$ and $A$. Then she measures $\sigma_{x}$ on qubit $a$ and sends
the result to Bob.

(2) Bob applies $I_{2}$ or $\sigma_{z}$ on qubit $b$ according to Alice's
measurement result. They can obtain the state%
\begin{equation}
\left\vert \Psi_{bAB}\right\rangle =\cos\left(  \alpha/2\right)  \left\vert
0\right\rangle _{b}\otimes\left\vert \Phi_{AB}\right\rangle +i\sin\left(
\alpha/2\right)  \left\vert 1\right\rangle _{b}\otimes\sigma_{z}^{A}\left\vert
\Phi_{AB}\right\rangle , \label{4}%
\end{equation}
where $\left\vert \Phi_{AB}\right\rangle $ is the initial state of the qubit
$A$ and $B$ and we discard the qubit $a$.

(3) Bob implements a controlled-phase gate
\begin{equation}
U_{bB}=\left\vert 0\right\rangle _{b}\left\langle 0\right\vert _{b}\otimes
I_{2}+\left\vert 1\right\rangle _{b}\left\langle 1\right\vert _{b}%
\otimes\sigma_{z}^{B} \label{5}%
\end{equation}
on qubit $b$ and $B$. They obtain the state%
\begin{equation}
\left\vert \Psi_{bAB}^{\prime}\right\rangle =\cos\left(  \alpha/2\right)
\left\vert 0\right\rangle _{b}\otimes\left\vert \Phi_{AB}\right\rangle
+i\sin\left(  \alpha/2\right)  \left\vert 1\right\rangle _{b}\otimes\sigma
_{z}^{A}\sigma_{z}^{B}\left\vert \Phi_{AB}\right\rangle . \label{6}%
\end{equation}

(4) Bob makes a positive operator-valued measure (POVM) on qubit $b$ with
three positive operators%
\begin{align}
E_{1}  &  =x\left\vert \Phi_{1}\right\rangle \left\langle \Phi_{1}\right\vert
,\left\vert \Phi_{1}\right\rangle =\frac{\cos\left(  \theta/2\right)  }%
{\cos\left(  \alpha/2\right)  }\left\vert 0\right\rangle +\frac{\sin\left(
\theta/2\right)  }{\sin\left(  \alpha/2\right)  }\left\vert 1\right\rangle
,\label{7}\\
E_{2}  &  =y\left\vert \Phi_{2}\right\rangle \left\langle \Phi_{2}\right\vert
,\left\vert \Phi_{2}\right\rangle =\frac{\sin\left(  \theta/2\right)  }%
{\cos\left(  \alpha/2\right)  }\left\vert 0\right\rangle -\frac{\cos\left(
\theta/2\right)  }{\sin\left(  \alpha/2\right)  }\left\vert 1\right\rangle
,\nonumber\\
E_{3}  &  =I_{2}-E_{1}-E_{2}.\nonumber
\end{align}
When Bob gets the result $E_{1}$, the desired gate $U\left(  \theta\right)  $
is implemented on the qubit $A$ and $B$. When Bob gets the result $E_{2},$ the
gate $-i\sigma_{z}^{A}\sigma_{z}^{B}U\left(  \theta\right)  $ is implemented
on the qubit $A$ and $B$, which can be changed into $U\left(  \theta\right)  $
after Bob tells Alice his measurement result. So when Bob gets the result
$E_{1}$ and $E_{2}$, they can succeed to implement the desired gate $U\left(
\theta\right)  $ on the qubit $A$ and $B$.

Our purpose is to maximize the successful probability%
\begin{equation}
p=\left\langle \Psi_{bAB}^{\prime}\right\vert E_{1}\left\vert \Psi
_{bAB}^{\prime}\right\rangle +\left\langle \Psi_{bAB}^{\prime}\right\vert
E_{2}\left\vert \Psi_{bAB}^{\prime}\right\rangle =x+y \label{8}%
\end{equation}
under the condition that $E_{1}$, $E_{2}$, and $E_{3}$ are positive operators,
which requires that%
\begin{equation}
x\geq0,y\geq0,TrE_{3}\geq0,DetE_{3}\geq0. \label{9}%
\end{equation}
Simple calculation shows that
\begin{equation}
TrE_{3}=2-x\left(  \frac{\cos^{2}\frac{\theta}{2}}{\cos^{2}\frac{\alpha}{2}%
}+\frac{\sin^{2}\frac{\theta}{2}}{\sin^{2}\frac{\alpha}{2}}\right)  -y\left(
\frac{\sin^{2}\frac{\theta}{2}}{\cos^{2}\frac{\alpha}{2}}+\frac{\cos^{2}%
\frac{\theta}{2}}{\sin^{2}\frac{\alpha}{2}}\right)  , \label{10}%
\end{equation}%
\begin{align}
DetE_{3}  &  =TrE_{3}-1+4xy/\sin^{2}\alpha\label{11}\\
&  =\frac{4}{\sin^{2}\alpha}\left[  \left(  x-\frac{1+\cos\theta\cos\alpha}%
{2}\right)  \left(  y-\frac{1-\cos\theta\cos\alpha}{2}\right)  -\frac{\cos
^{2}\alpha\sin^{2}\theta}{4}\right]  .\nonumber
\end{align}
Observe that the curve $DetE_{3}=0$ is a hyperbola and $TrE_{3}=0$ and
$DetE_{3}=0$ have no common point, the permissible $(x,y)$ is in the common
part of $x\geq0$, $y\geq0,$and the left lower part of $DetE_{3}\geq0$. This
part is concave and the maximal permissible $x$ is $\frac{\sin^{2}\alpha
}{2\left(  1-\cos\theta\cos\alpha\right)  }.$ We denote%
\begin{align}
\Delta &  =\left(  \frac{1+\cos\theta\cos\alpha}{2}-\sqrt{\frac{\cos^{2}%
\alpha\sin^{2}\theta}{4}}\right)  -\frac{\sin^{2}\alpha}{2\left(  1-\cos
\theta\cos\alpha\right)  }\label{12}\\
&  =\frac{\cos\alpha\sin\theta\left[  \cos\alpha\left(  \sin\theta+\cos
\theta\right)  -1\right]  }{2\left(  1-\cos\theta\cos\alpha\right)
}.\nonumber
\end{align}

(i) When $\Delta\leq0$, \textit{i.e.}, $\cos\alpha\left(  \sin\theta
+\cos\theta\right)  \leq1$, the successful probability $p$ achieve its maximum
at
\begin{equation}
(x,y)=(\frac{1+\cos\theta\cos\alpha-\sin\theta\cos\alpha}{2},\frac
{1-\cos\theta\cos\alpha-\sin\theta\cos\alpha}{2}), \label{13}%
\end{equation}%
\begin{equation}
p_{\max}=x+y=1-\sin\theta\cos\alpha. \label{14}%
\end{equation}

(ii) When $\Delta\geq0$, \textit{i.e.}, $\cos\alpha\left(  \sin\theta
+\cos\theta\right)  \geq1$, the successful probability $p$ achieve its maximum
at%
\begin{equation}
(x,y)=\left(  \frac{\sin^{2}\alpha}{2\left(  1-\cos\theta\cos\alpha\right)
},0\right)  , \label{15}%
\end{equation}%
\begin{equation}
p_{\max}=x+y=\frac{\sin^{2}\alpha}{2\left(  1-\cos\theta\cos\alpha\right)  }.
\label{16}%
\end{equation}

We give some remarks about our result. When we fix the rotation angle $\theta$
we observe how the maximal successful probability varies with $\alpha$
(entanglement). Notice that $\sin\theta+\cos\theta\geq1$, when $\alpha$ is
close with zero $\cos\alpha\left(  \sin\theta+\cos\theta\right)  \geq1,$ we
have%
\begin{equation}
p_{\max}=\frac{\sin^{2}\alpha}{2\left(  1-\cos\theta\cos\alpha\right)  },
\label{17}%
\end{equation}
which will increase as we increase $\alpha$. When $\alpha$ is close with
$\pi/2$ $\cos\alpha\left(  \sin\theta+\cos\theta\right)  \leq1,$ we have%
\begin{equation}
p_{\max}=1-\sin\theta\cos\alpha, \label{18}%
\end{equation}
which will increase as we increase $\alpha$, and $p_{\max}$ reaches unit when
the state we consumed is maximally entangled ($\alpha=\pi/2$).

When we fix the entanglement ($\alpha$), the situation is a little different.
When $\alpha\geq\pi/4$, we always have $\cos\alpha\left(  \sin\theta
+\cos\theta\right)  \leq1$ for any $\theta\in\left(  0,\pi/2\right]  $, so we
have $p_{\max}=1-\sin\theta\cos\alpha$ for any $\theta\in\left(
0,\pi/2\right]  $. This maximal successful probability will increase as we
decrease $\theta$ and when $\theta$ trends to zero $p_{\max}$ trends to unit.
When $\alpha<\pi/4$ we can divide the interval $\theta\in\left(
0,\pi/2\right]  $ into three small intervals to analyze our result. (i)
$\theta$ is close with zero and $\cos\alpha\left(  \sin\theta+\cos
\theta\right)  \leq1$. In this interval $p_{\max}=1-\sin\theta\cos\alpha$.
(ii) $\theta$ is close with $\pi/4$ and $\cos\alpha\left(  \sin\theta
+\cos\theta\right)  \geq1$. In this interval $p_{\max}=\frac{\sin^{2}\alpha
}{2\left(  1-\cos\theta\cos\alpha\right)  }$. (iii) $\theta$ is close with
$\pi/2$ and $\cos\alpha\left(  \sin\theta+\cos\theta\right)  \leq1$. In this
interval $p_{\max}=1-\sin\theta\cos\alpha$. In each small interval, the
maximal successful probability will increase as we decrease the rotation angle
$\theta$. In the first small interval when the rotation angle $\theta$ trends
to zero the probability $p_{\max}$ will trend to unit, and in the third small
interval when the rotation angle $\theta$ trends to its maximal value $\pi/2$
the probability $p_{\max}$ will trend to $1-\cos\alpha=2\sin^{2}\frac{\alpha
}{2}$, which is the maximal probability of transforming the entanglement we
consumed into a Bell state. In one word, the maximal successful probability we
achieve will increase as we increase the entanglement or decrease the rotation angle.

When the entangled state we use is not a Bell state, the successful
probability cannot be unit, but when we fail we know exactly what unitary gate
is implemented on our target qubits. We consider a simple case where when we
fail we use a Bell state to accomplish our original task including
compensating the "failing" unitary gate. Thus the controlled-rotation is
accomplished determinately. Our purpose is to find out what
controlled-rotations can be implemented determinately using entanglement
smaller than one ebit on average. We use Von Neuman entropy as the measurement
of entanglement since it is additive in a sense \cite{book}. The average
entanglement is%
\begin{align}
E\left(  \theta,\alpha\right)   &  =p_{\max}E_{\alpha}+(1-p_{\max
})(1+E_{\alpha})\label{19}\\
&  =1-p_{\max}+E_{\alpha},\nonumber
\end{align}
where
\[
E_{\alpha}=-\left(  \cos^{2}\frac{\alpha}{2}\right)  \log_{2}\left(  \cos
^{2}\frac{\alpha}{2}\right)  -\left(  \sin^{2}\frac{\alpha}{2}\right)
\log_{2}\left(  \sin^{2}\frac{\alpha}{2}\right)
\]
is the entanglement we use. The probability $p_{\max}$ is given in Eq. (17) or
(18) dependent on $\alpha$ is close with zero or close with $\pi/2$. We find
that about when $\theta<0.234\pi$ we can always find an $\alpha$ that makes
$E\left(  \theta,\alpha\right)  <1$, \textit{i.e.}, implementing
controlled-rotations using entanglement less than one ebit on average.

In summary, we give an efficient method to implement nonlocal gates using
entanglement. The successful probability trends to unit when the
controlled-rotation angle trends to zero or the entanglement we use trends to
a Bell state. Our method can also be used in remote implementation of quantum
operations \cite{huelqa,xiang}.

This work was funded by National Fundamental Research Program (Program No.
2001CB309300), National Natural Science Foundation of China (No. 10304017),
and Chinese Innovation Fund (No. Grant 60121503).


\begin{thebibliography}{99}                                                                                               %


\bibitem {bennett}C. H. Bennett, G. Brassard, C. Cr\'{e}peau, R. Jozsa, A.
Peres, and W. K. Wootters, Phys. Rev. Lett. \textbf{70}, 1895 (1993).

\bibitem {book}M. A. Nielsen and I. L. Chuang, \textit{Quantum Computation and
Quantum Information} (Cambridge University Press, Cambridge, 2000).

\bibitem {cirac}J. I. Cirac, W. D\"{u}r, B. Kraus, and M. Lewenstein, Phys.
Rev. Lett. \textbf{86}, 544 (2001).

\bibitem {dur}D. D\"{u}r, G. Vidal, and J. I. Cirac, Phys. Rev. Lett.
\textbf{89}, 057901 (2002).

\bibitem {vidal}G. Vidal and J. I. Cirac, Phys. Rev. Lett. \textbf{88}, 167903 (2002).

\bibitem {kraus}B. Kraus and J. I. Cirac, Phys. Rev. A \textbf{63}, 062309 (2001).

\bibitem {leifer}M. S. Leifer, L. Henderson, and N. Linden, Phys. Rev. A
\textbf{67}, 012306 (2003).

\bibitem {ye}M. Y. Ye, D. Sun, Y. S. Zhang, and G. C Guo, Phys. Rev. A
\textbf{70}, 022326 (2004).

\bibitem {eisert}J. Eisert, K. Jacobs, P. Papadopoulos, and M. B. Plenio,
Phys. Rev. A \textbf{62}, 052317 (2000).

\bibitem {reznik}B. Reznik, Y. Aharonov, and B. Groisman, Phys. Rev. A
\textbf{65}, 032312 (2002).

\bibitem {groisman}B. Groisman1 and B. Reznik, Phys. Rev. A \textbf{71},
032322 (2005).

\bibitem {zheng}Y. Z. Zheng, Y. Peng, and G. C. Guo, Chin. Phys. Lett.
\textbf{21}, 9 (2004).

\bibitem {chen}L. Chen and Y. X. Chen, Phys. Rev. A \textbf{71}, 054302 (2005).

\bibitem {duan}L. M. Duan and R. Raussendorf, Phys. Rev. Lett. \textbf{95},
080503 (2005).

\bibitem {huelqa}S. F. Huelga, M. B. Plenio, and J. A. Vaccaro, Phys. Rev. A
\textbf{65}, 042316 (2002).

\bibitem {xiang}G. Y. Xiang, J. Li, and G. C. Guo, Phys. Rev. A \textbf{71},
044304 (2005).
\end{thebibliography}
\end{document}